\documentclass{article}

\usepackage{arxiv}

\usepackage[utf8]{inputenc} 
\usepackage[T1]{fontenc}    
\usepackage{hyperref}       
\usepackage{url}            
\usepackage{booktabs}       
\usepackage{amsfonts}       
\usepackage{nicefrac}       
\usepackage{microtype}      
\usepackage{cite}
\usepackage{amsmath,amssymb,amsfonts} 
\usepackage{algorithmic}
\usepackage{graphicx}
\usepackage{textcomp}
\usepackage{xcolor,colortbl}
\usepackage{multirow}
\usepackage{tabularx}
\usepackage{float}
\usepackage{caption}
\definecolor{greenC}{rgb}{0.7,1,0.7}

\title{Speed-of-Sound Imaging using Diverging Waves}

\author{
Richard Rau,
Dieter Schweizer,
Valery Vishnevskiy, 
Orcun Goksel \\
Computer-assisted Applications in Medicine, 
ETH Zurich, Switzerland
}

\begin{document}
\maketitle

\begin{abstract}

Recent ultrasound imaging modalities based on ultrasound computed tomography  indicate a huge potential to detect pathologies is tissue due to altered biomechanical properties. 
Especially the imaging of speed-of-sound (SoS) distribution in tissue has shown clinical promise and thus gained increasing attention in the field -- with several methods proposed based on transmission mode tomography. 
SoS imaging using conventional ultrasound (US) systems would be convenient and easy for clinical translation, but this requires using conventional US probes with single-sides tissue access and thus pulse-echo imaging sequences. 
Recent pulse-echo SoS imaging methods rely on plane wave (PW) insonifications, which is prone to strong aberration effects for non-homogeneous tissue composition. 
In this paper we propose to use diverging waves (DW) for SoS imaging and thus substantially improve the reconstruction of SoS distributions.
We study this proposition by first plane wavefront aberrations compared to DW. 
We then present the sensitivity of both approaches to major parameterization choices on a set of simulated phantoms.
Using the optimum parameter combination for each method for a given transducer model and imaging sequence, we analyze the SoS imaging performance comparatively between the two approaches. 
Results indicate that using DW instead of PW, the reconstruction accuracy improves substantially, by over 22\% in reconstruction error (RMSE) and by 55\% in contrast (CNR).
We also demonstrate improvements in SoS reconstructions from an actual US acquisition of a breast phantom with tumor- and cyst-representative inclusions, with high and low SoS contrast, respectively.

\end{abstract}

\section{Introduction}

Medical imaging based on ultrasound (US) is a standard imaging modality in clinics thanks to the low cost, non-ionizing nature, portability and real-time imaging capability of conventional US systems.
In such systems, US is typically used to image tissue structures based on echo amplitude indicating local tissue reflectivity (B-Mode imaging), frequency shift indicating blood flow characteristics (Doppler imaging), or shear-wave propagation speed indicating elasticity properties~\cite{szabo_diagnostic_2013}. 
Additionally, specialized US imaging devices have emerged that use transmission based computed tomography (CT) approaches to characterize additional tissue properties such as speed-of-sound (SoS) and attenuation from arrival time and power-loss computation~\cite{mamou_quantitative_2013,malik_breast_2019}.
Especially for breast cancer, it was shown that a quantitative assessment of SoS bears tremendous potential, because it relates to the biomechanical properties of tissue, mainly the bulk modulus, and thus may act as an imaging biomarker for pathologies~\cite{li_vivo_2009,pratt_sound-speed_2007,li_breast_2017,malik_breast_2019,malik_quantitative_2018,johnson_noninvasive_2007}.
In comparison to shear-wave elastography, SoS was found to lead to a better \textit{ex vivo} tissue differentiation~\cite{glozman_method_2010} with high specificity for benign and malignant tumors~\cite{goss_compilation_1980,goss_comprehensive_1978,li_vivo_2009}. 
However, to image the SoS distribution in transmission-mode, double-sided access to tissue suspended in a water bath is required such as with two opposing transducers~\cite{malik_quantitative_2018}, ring shaped~\cite{duric_detection_2007} or full 3D~\cite{gemmeke_3d_2007} transducer geometries.
These are implementable only in costly and non-portable systems, require a technician to operate, and allow the imaging of only submersible body parts e.g.\ the breast and the extremities. 
A translatable application of SoS imaging to widely available conventional US systems thus necessitate the development and improvement of pulse-echo based SoS reconstruction techniques. 

Recently several methods have been proposed for SoS imaging using conventional linear array transducers used in the clinics.
For instance in~\cite{sanabria_hand-held_2016,sanabria_speed--sound_2018}, it was proposed to use a limited-angle computed tomographic reconstruction for SoS imaging by recording the time-of-flight values of reflections from passive reflector placed at a known distance from the transducer.
Obviating the need for a reflector, small misalignments between images acquired at different plane-wave angles were used in~\cite{jaeger_computed_2015} to reconstruct SoS distribution using a Fourier domain reconstruction approach. 
In~\cite{sanabria_spatial_2018}, SoS reconstruction in the spatial domain was shown to yield improved accuracy and less artifacts. 
In~\cite{stahli_forward_2019} it was proposed for PW transmits to adapt receive apertures dynamically when beamforming different image locations to minimize spatial point-spread function (PSF) variation, in order to improve displacement estimation used for SoS reconstruction. 

In clinical settings, several works have studied SoS imaging using transmission-mode and water-submerged systems, e.g.\ for breast tissue classification~\cite{klock_visual_2017}, solid mass differentiation~\cite{iuanow_accuracy_2017}, and imaging human-knee~\cite{wiskin_3d_2019}.
Using conventional transducer in pulse-echo mode, SoS has been studied clinically for quantifying muscle loss~\cite{sanabria_speed_2018_sarc} and breast density~\cite{sanabria_breast-density_2018}, as well as for differential diagnosis of breast cancer (invasive ductal carcinoma vs.\ fibroadenoma)~\cite{Ruby_breast_19}.

In addition to the clinical use of SoS imaging to characterize pathological changes of the biomechanical properties in tissue, the quantitative SoS information can furthermore help to compensate and correct for aberrations, and hence to improve any other US imaging modality. 
For instance, beamforming is based on delay calculation assuming a constant SoS, an incorrect assumption of which not only reduces B-mode resolution but may also affect following processing such as in motion estimation for shear-wave elastography. 
With the knowledge of SoS distribution, aberrations can be corrected as demonstrated in~\cite{rau_ultrasound_2019,jaeger_full_2015,ali_distributed_2018}.  

Despite many promising examples, pulse-echo SoS imaging using conventional transducers is still a challenging task to achieve robust reconstructions under changing conditions. 
In contrast to PW SoS imaging approaches, we herein propose a transmit sequence with diverging waves (DW) to yield reduced aberration artifacts and thus improved reconstructions. 
We study its feasibility for SoS imaging comparatively to PW, also given the effect of PSF centering via adapted receive aperture.

For any reconstruction method, typically several algorithmic parameters need to be tuned and the settings for one image may not be ideal for another.
Consequently the comparison and validation of proposed new methods may also be affected, since an algorithmic choice improving one image may degrade another.
To alleviate such image dependence and to demonstrate generalizability independent of a chosen setting, we analyze the compared methods for a set of several simulated images with varied SoS heterogeneities, and present evaluation results based on average metrics.

\section{Methods}

Below we first motivate the use of diverging waves for reduced aberration artifacts using an example heterogeneous case.
We then summarize the methodological steps used for SoS reconstruction, before presenting an experimental evaluation.

\subsection{Aberration Effects with Plane and Diverging Waves}

\begin{figure}
  \centering
    \includegraphics[width=.98\textwidth]{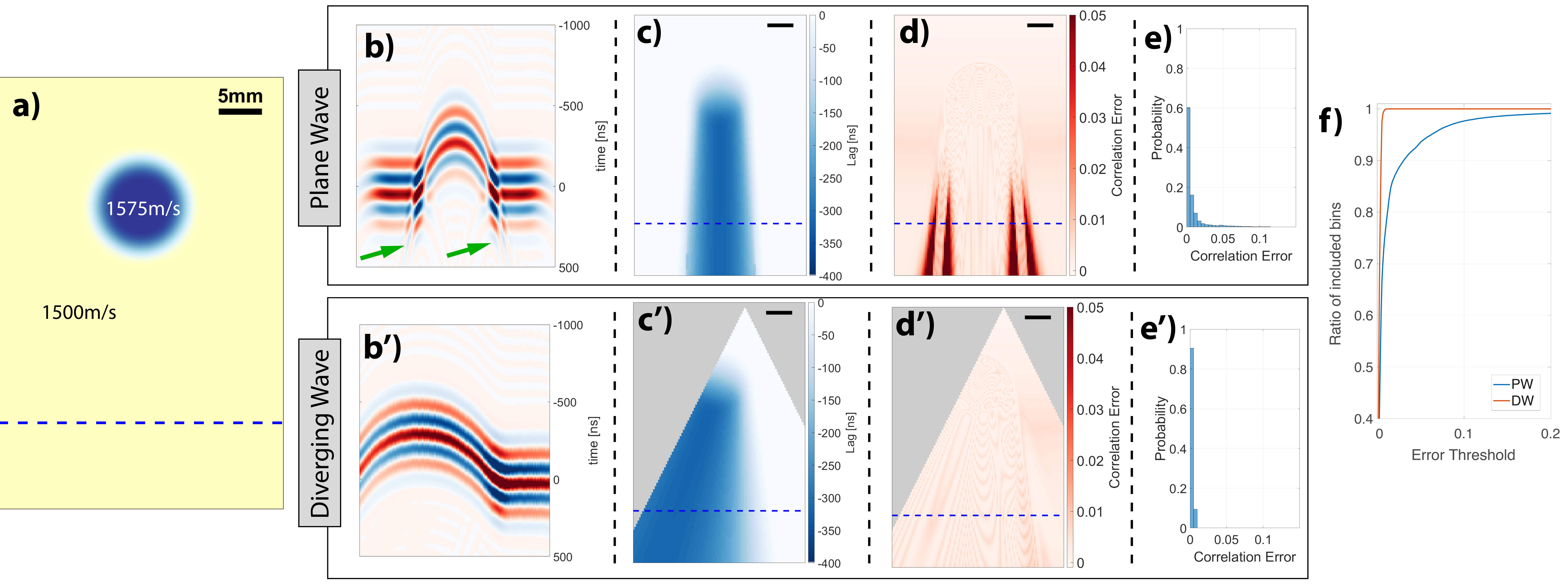}
\captionof{figure}{
Wavefront aberration comparisons in simulations using acoustic recordings in the entire imaging field. 
(a)~Heterogeneous SoS distribution with a 10\,~mm circular inclusion of 5\% contrast (i.e., 1575m/s on a 1500m/s substrate).
The SoS map was Gaussian-smoothed (standard deviation of 1\,mm) to avoid sharp edges less typical in human tissues. 
The wavefronts arriving at the dashed line in~(a) are shown (b)~for $0^\circ$ plane-wave transmit and (b')~for the diverging wave from Tx element 81. 
The time on the y-axis in (b,b') is referenced to the arrival time of the wavefront peak for a homogeneous simulation without the SoS inclusion. 
The green arrows indicate the aberration effects due to diffraction.
(c/c')~Lags obtained by cross correlating the signals with the signals from the homogeneous setting.
(d/d')~Correlation error (i.e.\ $1-$correlation coefficient) obtained from normalized cross correlation.
(e/e')~Probability distribution of the correlation error based on three \{-10,0,10\}$^\circ$ PW and 32 Tx-element DW datasets.
(f)~Cumulative distribution function from (e/e').
}
\label{fig:hypo}
\end{figure}

In order to motivate the use of diverging waves, we first demonstrate the aberration effects comparatively to plane waves, on a synthetic simulation example, seen in Fig.~\ref{fig:hypo}.
Using the MATLAB toolbox k-wave\cite{treeby_k-wave:_2010}, we simulate different transmit schemes and record the spatio-temporal acoustic signal at each and every point in the entire imaging field-of-view (FOV).
Note that this hypothetical recording is not possible in actual tissue, where we only receive the returned echos, and can only be approximated with a water-bath hydrophone experiment.
For both transmit settings, we run two simulations: one with an SoS inclusion and one for a homogeneous case with no SoS inclusions, in order to comparatively quantify the affect of aberrations introduced by the inclusion

Consider the wavefronts arriving at a certain depth (e.g., marked with the dashed line in Fig.\ref{fig:hypo}a).
As expected, the wavefronts passing through the inclusion would arrive earlier at such depth, compared to a no inclusion scenario, cf.\,Fig.\,\ref{fig:hypo}b/b'. 
Besides such earlier arrival, one can observe the strong aberration effects below the edges of the inclusion for the PW case (shown with the arrows in Fig.\,\ref{fig:hypo}b), mainly caused by diffraction. 
Such aberrations could aggravate when the echos are considered, and would largely hinder any post-processing such as delay estimation for SoS reconstruction.
To demonstrate this, we perform here a delay estimation only for the transmit side using normalized cross-correlation (NCC) between spatio-temporal data at all image points for with-inclusion and no-inclusion cases.
Fig.\,\ref{fig:hypo}c/c' show the estimated delays and Fig.\,\ref{fig:hypo}d/d' the correlation error in delay estimation, the sub-figures for PW and Dw, respectively.
As seen in Fig.\,\ref{fig:hypo}d, delay estimation accuracy is quite low with PW tranmits, compared to DW. 
This can be more generally stated via statistics from multiple PW and DW settings, shown in Fig.\ref{fig:hypo}e/e' as probability density functions, where the PW case is seen to have errors further away from zero.
To better illustrate this, the cumulative distribution functions of plotted in Fig.\ref{fig:hypo}f, which 
indicates the number of highly aberrated readings that forestall accurate displacement estimation.
As can be seen, given any NCC tolerance/threshold, DW would yield much superior displacement estimation than PW, for aberrations typical to expect in in-vivo tissue.
For instance, for an NCC tolerance of minimum 0.99, 20\% of PW readings would be below this threshold, while all DW readings would be within bounds -- which is a large margin considering the single small inclusion given a large homogeneous FOV.

In the context of SoS imaging, such wavefront distortions lead to incoherent signal summations in receive beamforming, thus corrupting displacement estimations, which in turn potentially degrade the SoS reconstruction.
By reducing these wavefront aberrations, DW can yield improved SoS imaging, which is studied later in our experiments.

\subsection{Speed-of-Sound Imaging}

\begin{figure}
  \centering
    \includegraphics[width=.98\textwidth]{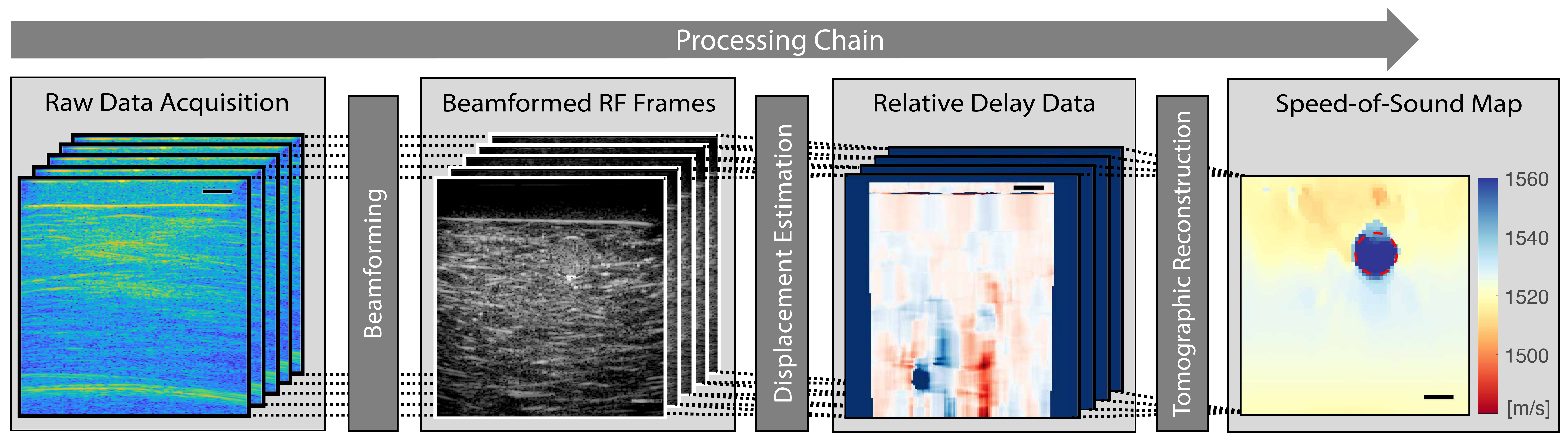}
\captionof{figure}{Processing chain for SoS imaging: 
Raw channel data acquired with different Tx sequences are first beamformed. 
Then, apparent displacements are computed between several beamformed RF frames, yielding (potentially noisy estimates of) relative delay maps.
Based on the utilized Tx sequence and the respective Tx-Rx wave paths, the forward problem of relative delays is formulated as a linear system.
The inverse problem of tomographic reconstruction is then solved using an iterative optimization algorithm to reconstruct the SoS distribution in the tissue.
}
\label{fig:pipeline}
\end{figure}

We herein use a limited-angle computed tomography (LA-CT) reconstruction method in the spatial domain, similarly to~\cite{sanabria_spatial_2018} with the adaptations described below.
The fundamental imaging principle and an overview of the data processing is sketched out in Fig.~\ref{fig:pipeline}.
First, raw data is acquired based on a PW or DW transmit (Tx) sequence, which both involve multiple transmits with receive (Rx) on all element channels, i.e.\ full-matrix capture (FMC).
Then, separately for each Tx, these Rx signals are beamformed into spatial RF frames, between which apparent local displacements are estimated to be next used to reconstruct an SoS map.

\subsubsection{Beamforming} \label{sec:BF}

To beamform with the received raw channel data from the PW or DW transmit sequences (cf.\,Fig.~\ref{fig:pipeline}), we herein employ a conventional delay-and-sum algorithm.
The delays are computed with an assumed constant SoS of $1500\,\mathrm{m/s}$ for the simulations.
For both transmit schemes and all RF frames, beamforming is performed on a fixed Cartesian grid aligned with the transducer surface, for a fixed sampling space for the subsequent displacement estimation between these frames.

We present results with two different beamforming choices: with \textit{full} Rx aperture and with an \textit{adapted} Rx aperture. 
In the full Rx aperture case signals from all channels are fed into beamforming, while still subjected to a dynamic aperture per imaging depth (herein with F-number = $1$), which results in Rx aperture staying centered above each beamformed image point. 
As Tx arrival directions to a point keep changing with each Tx, this then yields a PSF varying between different transmits, impeding the subsequent displacement estimation.
This is remedied with an adapted Rx aperture (Fig.\,\ref{fig:acqscheme}a), which is centered for each beamformed image point such that the PSF between Tx events to be displacement estimated are aligned as described in \cite{stahli_forward_2019}. 
Depending on the RX aperture chosen, the PSFs can be aligned at different angles for the same Tx event.
Each Tx event here is beamformed with three PSF angle alignments: $\psi_{\mathrm{psf}}=0^\circ$ and $\pm15^\circ$.
For all transmits, we utilize a fixed Cartesian beamforming grid of $N_x\times N_z$.

\subsubsection{Displacement Estimation} \label{sec:DT}

\begin{figure}
  \centering
    \includegraphics[width=.98\textwidth]{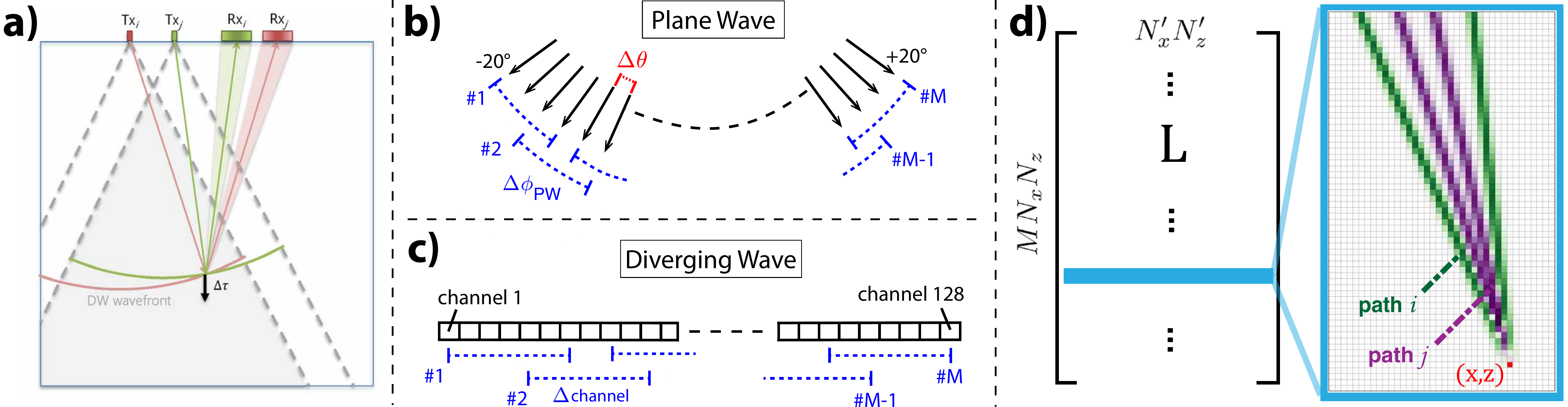}
\captionof{figure}{Illustration of PW and DW transmit schemes:
(a)~Sketch of PW acquisition with arrows indicating PW normals, i.e.\ the travel direction.
The displacement tracking from the acquired images is carried out between angle differences of $\Delta\theta$.
The relative delay data is obtained by accumulating the displacement estimation to an angular disparity $\Delta\phi_{PW}$. 
In total $M$ relative delay estimates are used in the reconstruction. 
(b)~Sketch of the parameters used for the diverging wave based reconstruction. 
The DWs are created with a single channel of the $N_c=128$ channel transducer.
For both cases, PW and DW, the mid angles / mid channels of each combinations are equally distributed over the angle / channel range, i.e. between $\pm20^\circ$ or between channels $[1,128]$.
(c)~Illustration of the Tx (determined by the $k$-vector) and Rx paths (determined by the receive aperture) for two DWs $i,j$ when an adapted RX aperture is chosen.
The displacement estimation in the axial direction yields the relative delay data $\Delta\tau$ in each beamformed pixel. 
(d)~A sample row of the differential path matrix \textbf{L} modeling the link between the SoS distribution and the relative delay data at pixel $(x,z)$.
The bipolar map describes positive values for path $i$ (green) and negative for path $j$ (purple).
In Rx only center path contributes to \textbf{L}, as also indicated in (c). 
}
\label{fig:acqscheme}
\end{figure}

For the apparent displacement estimation between beamformed RF frames (cf. Fig.~\ref{fig:pipeline}), we use a normalized cross-correlation algorithm in the axial direction, similarly to~\cite{sanabria_spatial_2018}.
For a constant computational complexity and to keep the data input into the reconstruction constant, we herein compute the relative delay data for a fixed number of $M=9$ combinations, yielding an apparent displacement vector of $\boldsymbol{\Delta\tau}$$\in$$\mathbb{R}^{M N_x N_z}$. 

For the PW case, the relative delays from an angle separation of $\Delta\phi$ (see Fig.~\ref{fig:acqscheme}b) are used in the reconstruction forward problem. 
Nevertheless, the actual displacement estimations are performed using cross-correlation between PW angles with a smaller increment $\Delta\theta$, in order to prevent speckle decorrelation and artefactual readings due to phase wrapping. 
To obtain the relative delay data for the larger disparities $\Delta\phi$, the delay readings from the $\Delta\theta$ increments are then accumulated. 
As the choice of $\Delta\phi$ highly affects SoS reconstructions, we varied these parameters to find the optimal setting, as later demonstrated in our results in Section~\ref{sec:sossim}. 
In the literature on pulse-echo SoS imaging, delay accumulations were performed for increments of $\Delta\theta = 0.5^\circ$ \cite{sanabria_speed--sound_2018,jaeger_computed_2015} or $\Delta\theta = 2^\circ$ \cite{stahli_forward_2019}, both settings of which we studied in our experiments later below. 

For the DW case, the relative delays are obtained by a frame selection as illustrated in Fig.~\ref{fig:acqscheme}c. 
Here, the delays are directly estimated using cross-correlation based displacement tracking between consecutive single element transmissions separated by $\Delta\mathrm{channel}$.
As can be expected, this setting highly affects the quality of the relative delays measurements:
On the one hand, for a small element separation, e.g. using consecutive channels, apparent displacements can be below the tracking noise level, as also illustrated later in our experiments.  
On the other hand, for large element separations, coherent speckle pattern changes drastically, precluding displacement estimation. 
Given this tradeoff, an optimal element separation $\Delta\mathrm{channel}$ is expected, which is studied later below for our experimental setup.

For the reconstruction, the projection of estimated displacements in the corresponding Tx-Rx direction is used, i.e.\ by multiplying them by $\cos(\psi_{\mathrm{psf}})$.

\subsubsection{Speed-of-Sound Rconstruction}

To reconstruct the SoS distribution based on the relative delay data (cf. Fig.~\ref{fig:pipeline}), an inverse problem is formulated to reconstruct the slowness $\boldsymbol{\hat\sigma}\in\mathbb{R}^{N_x'N_z'}$ on a $N_x'\times N_z'$ spatial grid, which is just the inverse of the SoS:
\begin{equation}\label{eq:sosrecon}
 \boldsymbol{\hat\sigma} = \arg \min_{\boldsymbol{\sigma}}
 \| \textbf{L}(\boldsymbol{\sigma-\sigma_0}) - \boldsymbol{\Delta\tau} \|_1  +  \lambda \|\textbf{D}\boldsymbol{\sigma} \|_1\ \ 
\end{equation}
The differential path matrix $\textbf{L}$$\in$$\mathbb{R}^{M N_x N_z\times N_x' N_z'}$ here links the relative slowness distribution $\boldsymbol{\sigma-\sigma_0}$ to the relative delay measurements; for instance, in Fig.\,\ref{fig:acqscheme}d the delay measurement at pixel $(x,z)$ is sensitive to SoS variation along the illustrated paths between the beamformed images $j$ and $i$.
The $\boldsymbol{\sigma_0}$ describes the initial slowness, which was used to compute the delays of the beamformed RF data to compute the delays. 

The regularization matrix \textbf{D} together with the weight $\lambda$ controls the amount of spatial smoothness and is essential due to ill-conditioning of $\textbf{L}$. 
\textbf{D} implements LA-CT specific image filtering aimed to suppress streaking artifacts along wave propagation directions via anisotropic weighting of horizontal, vertical and diagonal gradients. 
For the corresponding directions either a Sobel (horizontal and vertical) or a Roberts kernel (diagonal) is used as well as an $\kappa = 0.9$ anisotropic weighting, similar to~\cite{sanabria_spatial_2018}. 

For computational efficiency, we restrict the number of relative delay data readings $\boldsymbol{\Delta\tau}$ in eq.~(\ref{eq:sosrecon}) to $10^4$, which are randomly selected from all $\mathbb{R}^{M N_x N_z}$ recordings.

\section{Materials and Experiments}

\subsection{Simulation Experiment} \label{sec:simSOS}

To evaluate how accurate SoS heterogeneities can be imaged using the above explained SoS imaging method based on PWs or DWs, we simulated a pulse-echo scenario, where a linear transducer is simulated and the echos at each element are recorded.
In total 28 SoS heterogeneity cases are simulated (see first rows in Fig.\ref{fig:sim_recons}a/b), which are divided into two subsets.

The first subset (cases 1-6 and 28) is composed of seven defined shapes on a homogeneous background substrate of $1500\,\mathrm{m/s}$. 
The inclusions have a SoS contrast of either $-2\%$ (i.e. $1470\,\mathrm{m/s}$) or $+2\%$ (i.e. $1530\,\mathrm{m/s}$) and are elliptically shaped in cases 1-6 and circular shaped in the last case (28). 
In the last case, we simulate a setting with two inclusions.

The second subset (cases 7 - 27) is composed of randomly shaped inclusions with SoS values between [$1450,1550]\,\mathrm{m/s}$.
The substrate SoS values of these randomly shaped inclusions are varied in two ways:
1) The average SoS of the substrate is no longer fixed to  $1500\,\mathrm{m/s}$, but take values between [$1485,1515]\,\mathrm{m/s}$. 2) Each substrate is varied locally between $\pm3\,\mathrm{m/s}$.
Such substrate variations are important to evaluate how reconstructions would perform with natural tissue variation.
To allow for displacement estimation (cf. Fig.~\ref{fig:pipeline}), a fully-developed speckle pattern is required, realized herein by increasing a random 10\% set of the medium pixels by a slight perturbation in density value.

A linear array transducer is modeled with $N_c = 128$ channels and a $300\,\mu\mathrm{m}$ pitch. 
We used transmit pulses of $f_c=5\,\mathrm{MHz}$ center frequency with 3 half cycles. 
All simulations (including in Fig.~\ref{fig:hypo}) were run with a spatial discretization of $75\mu$m pixels and a temporal resolution of $6.25\,\mathrm{ns}$ (i.e., $160\,\mathrm{MHz}$ sampling frequency) to allow for an accurate sampling of the wave propagation.
Herein, for each case a full-matrix capture with multi-static transmission was simulated first and then recomposed into corresponding PWs or DWs using synthetic aperture Tx/Rx beamforming. 
To reduce high frequency artifacts, we additionally applied a 60\% band-pass filter on the channel data. 

For each simulation setup, 81 PW angles were computed (ranging between -20$^\circ$ and 20$^\circ$ with a step size of 0.5$^\circ$ and Tukey apodization) and for the DW we used single element transmissions (see Fig.~\ref{fig:acqscheme}a/b).
All raw data was beamformed based on a constant SoS assumption of $1500\mathrm{m/s}$.

\subsection{Phantom Experiment}

For data acquisition of the breast phantom (CIRS Multi-Modality Breast Biopsy and Sonographic Trainer, Model 073, CIRS Inc., Norfolk, VA, USA), we recorded unbeamformed RF data using a UF-760AG ultrasound system (Fukuda Denshi, Tokyo, Japan) with FUT-LA385-12P linear array transducer ($N_c = 128$ channels, $300\,\mu\mathrm{m}$ pitch). 
Data was acquired with $f_c=5\,\mathrm{MHz}$ center frequency and 4 half cycles pulses.
To increase the signal-to-noise ratio, data was transmitted using Wals-Hadamard coded pulses~\cite{villaverde_ultrasonic_2016} with subsequent decomposition into angled PWs or DWs, respectively. 
All raw data was beamformed based on a constant SoS assumption of $1470\mathrm{m/s}$, which is approximately the nominal SoS value of the main phantom material.

\subsection{Evaluation Metrics}

For a quantitative analysis of the SoS reconstruction $\hat{\boldsymbol{c}}=1/\hat{\boldsymbol{\sigma}}$ in simulation, we used the following metrics:
\begin{itemize}
    \item Root-mean-squared-error: 
    $\mathrm{RMSE} =\sqrt{\|\hat{\boldsymbol{c}}-\boldsymbol{c}^\star\|_2^2 / N}$.
    \item Contrast-to-noise ratio: $\mathrm{CNR} = 2(\mu_{\mathrm{inc}}-\mu_{\mathrm{bkg}})^2/\sqrt{\sigma_{\mathrm{inc}}^2 + \sigma_{\mathrm{bkg}}^2}$, mean value $\mu$ and variance $\sigma^2$. 
\end{itemize}
The subscripts $\cdot_\mathrm{inc}$ and $_{\mathrm{bkg}}$ denote the region of the inclusion and the background, respectively. 
Note that the $\mathrm{CNR}$ was only computed for cases 1-18, where the inclusion had an SoS contrast is at least 15\,m/s, i.e.\ 1$\%$ compared to the substrate.

\section{Results and Discussion}

\subsection{Simulations}\label{sec:sossim}

\begin{figure}
  \centering
    \includegraphics[width=.98\textwidth]{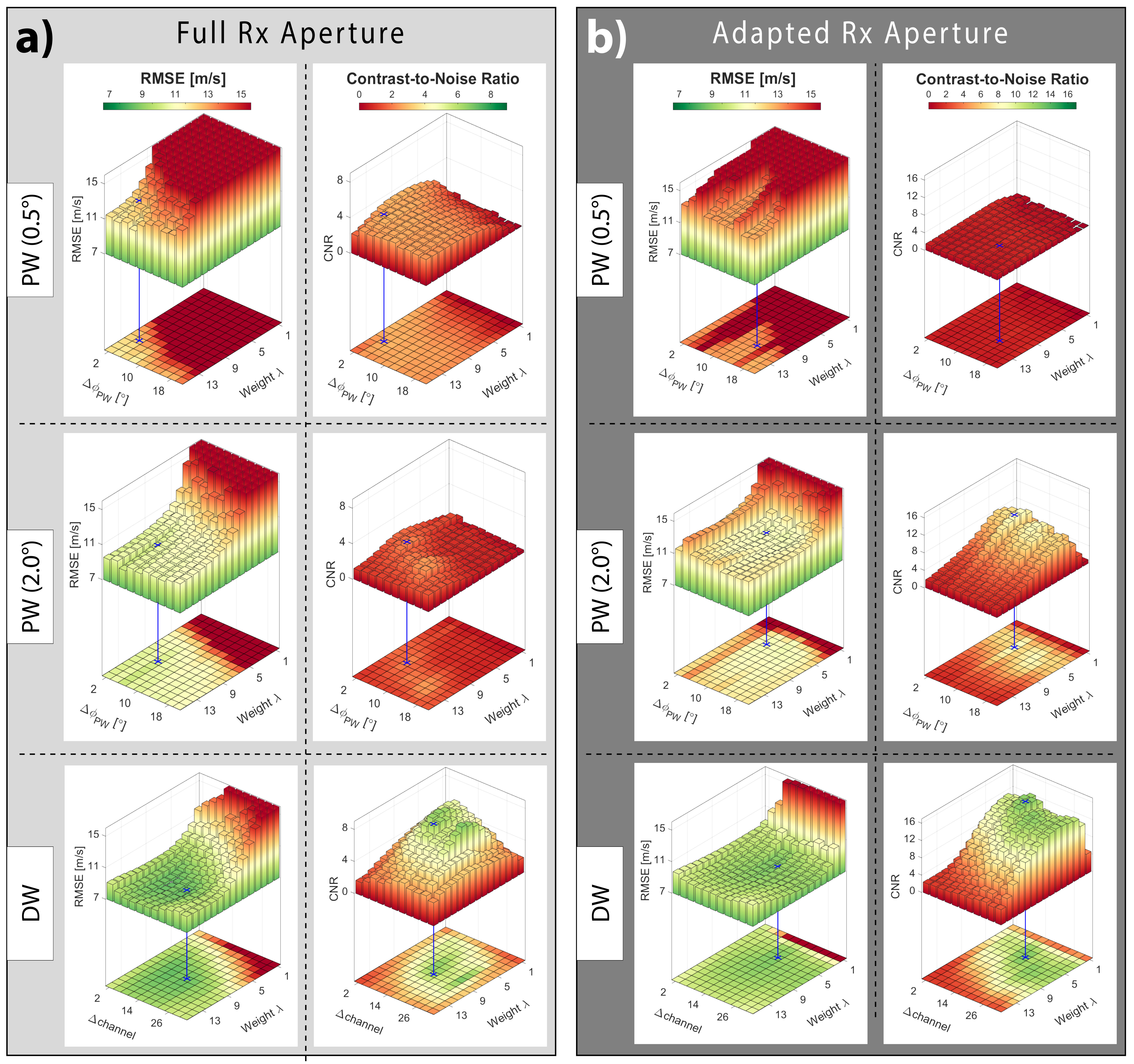}
\captionof{figure}{Evaluation of the RMSE and CNR with respect to the regularization weight, the parameter $\Delta\phi_{PW}$ for PW, and the parameter $\Delta\mathrm{channel}$ for DW.
Note that these two parameters act in a similar way in the two methods, as both describe the magnitude of speckle patter disparity.
(a)~Evaluation using full Rx aperture as in~\cite{jaeger_computed_2015,sanabria_spatial_2018}, and (b) using adapted Rx aperture as in~\cite{stahli_forward_2019,rau_ultrasound_2019}. Left columns show the $\mathrm{RMSE}$ for PW with $\Delta\theta = 0.5^\circ$ and  $\Delta\theta = 2.0^\circ$ and for the DW. 
The right columns present the $\mathrm{CNR}$ values.  
The blue bars indicate the optimal parameter values for each of the 6 methods, determined based on average ranking of the mean $\mathrm{RMSE}$ and $\mathrm{CNR}$ values displayed herein.
}
\label{fig:barplots}
\end{figure}

First, a sensitivity analysis with respect to major parametrization choices was performed for the corresponding transmission sequences (PW and DW).
We used a simulated phantom dataset of 28 ground-truth SoS distributions, representative of different characteristics in inclusion shape, size and SoS contrast as well as background SoS variations.
These datasets are then evaluated in terms of $\mathrm{RMSE}$ as well as in terms of $\mathrm{CNR}$, indicating how well the inclusions can be separated from the background, which is of major importance in the context tumor detection/characterization. 

The optimal parameters ($\Delta\phi$, $\Delta\mathrm{channel}$ and regularization weight $\lambda$) are then selected as follows: 
For each parameter combination, average $\mathrm{RMSE}$ and $\mathrm{CNR}$ across all 28 sample images was computed, as also plotted in Fig.~\ref{fig:barplots}. These values were then ranked from best to worst (i.e.\ first lowest for RMSE, and highest for CNR), and the optimal parameter set (cf.\,Table\,~\ref{tab:FinalResults}) was picked as the one minimizing the average rank of the two metrics.

The results are also summarized in table~\ref{tab:FinalResults}, where it can be seen that with DW the best results are achieved with an overall $\mathrm{RMSE}=7.3$ and $7.6$, respectively, for full and adapted Rx aperture cases. 
For the PW case, the best achievable results are at least $2$m/s on average poorer with $\mathrm{RMSE}=9.4$ and $10$, respectively.
The contrast is also substantially improved with DW to $\mathrm{CNR}=7.8$ and $16.1$, respectively, with over 35\% improvement compared to best case PW results of $\mathrm{CNR}=3.4$ and $10.3$.

\begin{figure}
  \centering
    \includegraphics[width=.98\textwidth]{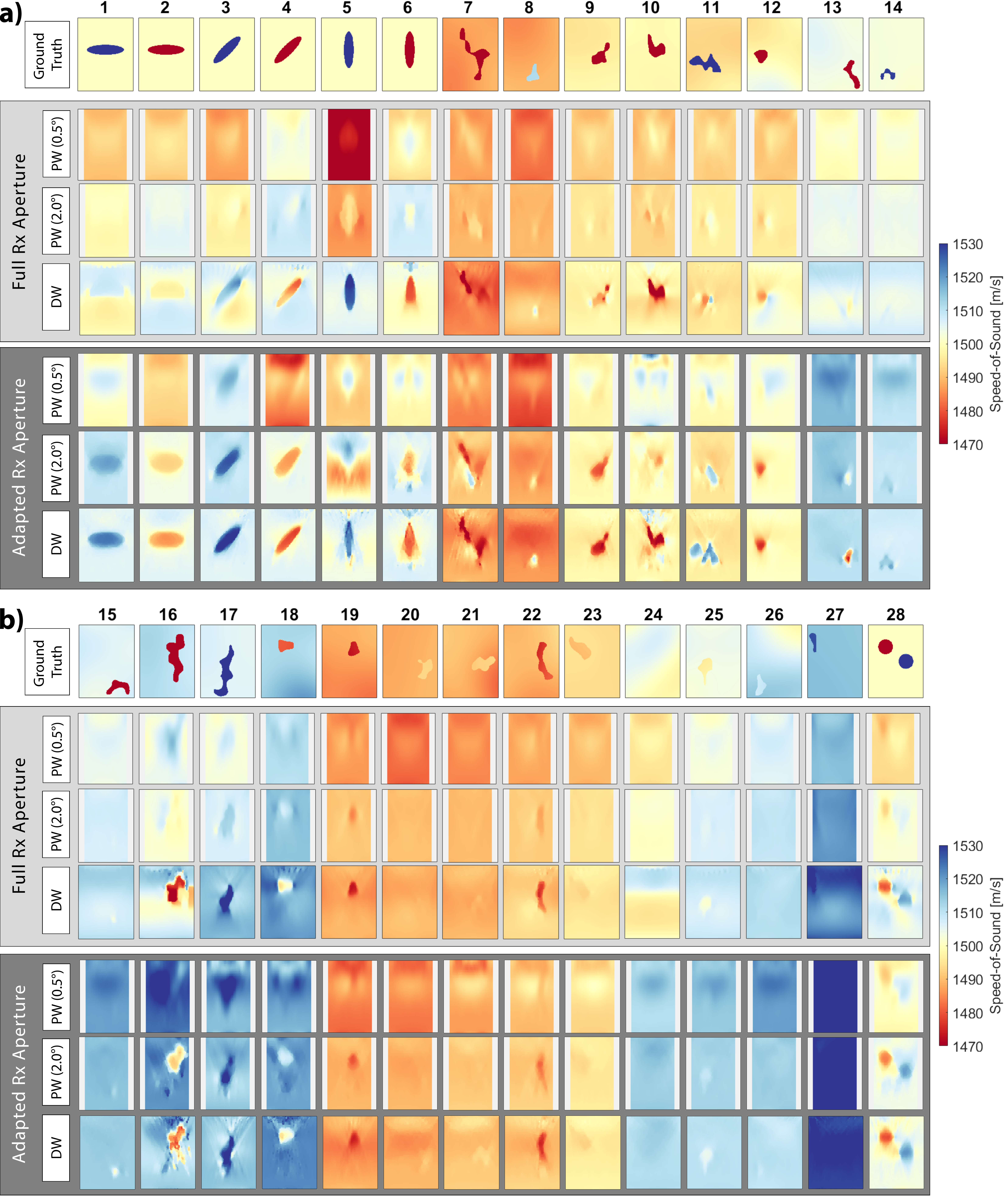}
\captionof{figure}{Reconstructions of the 28 numerical phantoms (a,b) using the optimal parameter settings given in Table~\ref{tab:FinalResults} as identified in Fig.\,\ref{fig:barplots}. 
The light grey boxes show reconstructions using full Rx aperture as in \cite{sanabria_spatial_2018,jaeger_computed_2015}, whereas the dark grey boxes using the adapted Rx apertures to align the PSF as in \cite{rau_ultrasound_2019,stahli_forward_2019}. 
Within each box, reconstructions with PW and the proposed DW are shown.
The dimension of each image is $38\,\mathrm{mm}\times50\,\mathrm{mm}$.
For PW reconstructions, a small margin of 10\% is masked out on both sides, since the angled PW apodization on the edges cause major artifacts, from beamforming to reconstruction.
}
\label{fig:sim_recons}
\end{figure}

Using the determined optimal parameter settings, the reconstructions of the 28 test images are shown in Fig.~\ref{fig:sim_recons}.
The DW-based SoS reconstructions are seen to be superior to PW-based in almost all cases. 
Especially with the case \#28 in Fig.~\ref{fig:sim_recons} with both higher and lower inclusions, DW is seen to perform significantly superior, regardless of the choice of aperture.

For the PW cases, the best reconstructions are achieved using an angle accumulation of $\Delta\theta=2^\circ$ and an adapted receive aperture, in agreement with~\cite{stahli_forward_2019}.
It is worthwhile to note that the adaptive receive aperture setting has a similar $\mathrm{RMSE}$ (of 10.0 vs.\ 9.4\,m/s) compared to the full receive aperture setting, leads to a substantially improved $\mathrm{CNR}$ (of 10.3 vs.\ 2.7). 
A similar trend is observed in the DW case with adapted vs.\ full receive aperture settings ($\mathrm{RMSE}$: 7.6 vs.\ 7.3\,m/s; $\mathrm{CNR}$: 16.1 vs.\ 7.8).
Notwithstanding the aperture differences, the overall SoS imaging is significantly improved using DW vs. PW.

Note that for PW we focus on the center part of the image and mask out 10\% on both sides of the imaging region (Fig.~\ref{fig:sim_recons}), since the apodization of the angled PW cause significant artifacts in these image regions, as was also discussed in \cite{Ruby_breast_19}.
Accordingly, $\mathrm{RMSE}$ and $\mathrm{CNR}$ were computed in these shown central regions. 
This then gives PW a slight advantage in evaluation, despite which we show that the proposed DW performs superior.

Beamforming was conducted assuming a constant $1500\mathrm{m/s}$, although the actual background SoS differed sometimes over $15,\mathrm{m/s}$. Despite deviations of the mean SoS values compared the assumed SoS in beamforming, SoS reconstructions are seen to still perform relatively well.
This is relevant to real case scenarios where the exact SoS values are not known a priori. 

\begin{table}[htbp]
  \centering
  \caption{Optimal parameter settings and the corresponding results}\vspace{3mm}
    \begin{tabular}{|l|r|r|r|r|r|r|}
    \multicolumn{1}{l}{} & \multicolumn{3}{c|}{\textbf{Full Receive Aperture}} & \multicolumn{3}{c}{\textbf{Adapted Receive Aperture}} \\
\cline{2-7}    \multicolumn{1}{l|}{\textbf{Settings}} & PW (0.5$^\circ$)    & PW (2.0$^\circ$)    & DW           & PW (0.5$^\circ$)    & PW (2.0$^\circ$)    & DW \\
    \hline
    $\Delta$          & 4$^\circ$  & 6$^\circ$  & 17 channels  &   14$^\circ$  & 8$^\circ$ & 17 channels \\
    \hline
    $\lambda$ [a.u.]  & 12  & 10  & 9            &   11   & 5  & 5 \\
		\hline
    \midrule
  \multicolumn{1}{l}{\textbf{Results}} &   \multicolumn{6}{r}{} \\
    \hline
    $\mathrm{RMSE}$ [m/s]   &  12.0        &  9.4         &  7.3         & 13.12        &     10.0       &    7.6      \\
    \hline
    $\mathrm{CNR}$ [a.u.]  & $3.4\pm 4.1$   &  $2.7\pm 3.8$ &   $7.8\pm 11.3$   & $1.8\pm2.3 $ & $10.3\pm9.9$    & $ 16.1\pm 12.8$  \\
    \hline
    \end{tabular}%
  \label{tab:FinalResults}%
\end{table}%

\subsection{Phantom Experiment}

\begin{figure}
  \centering
    \includegraphics[width=.7\textwidth]{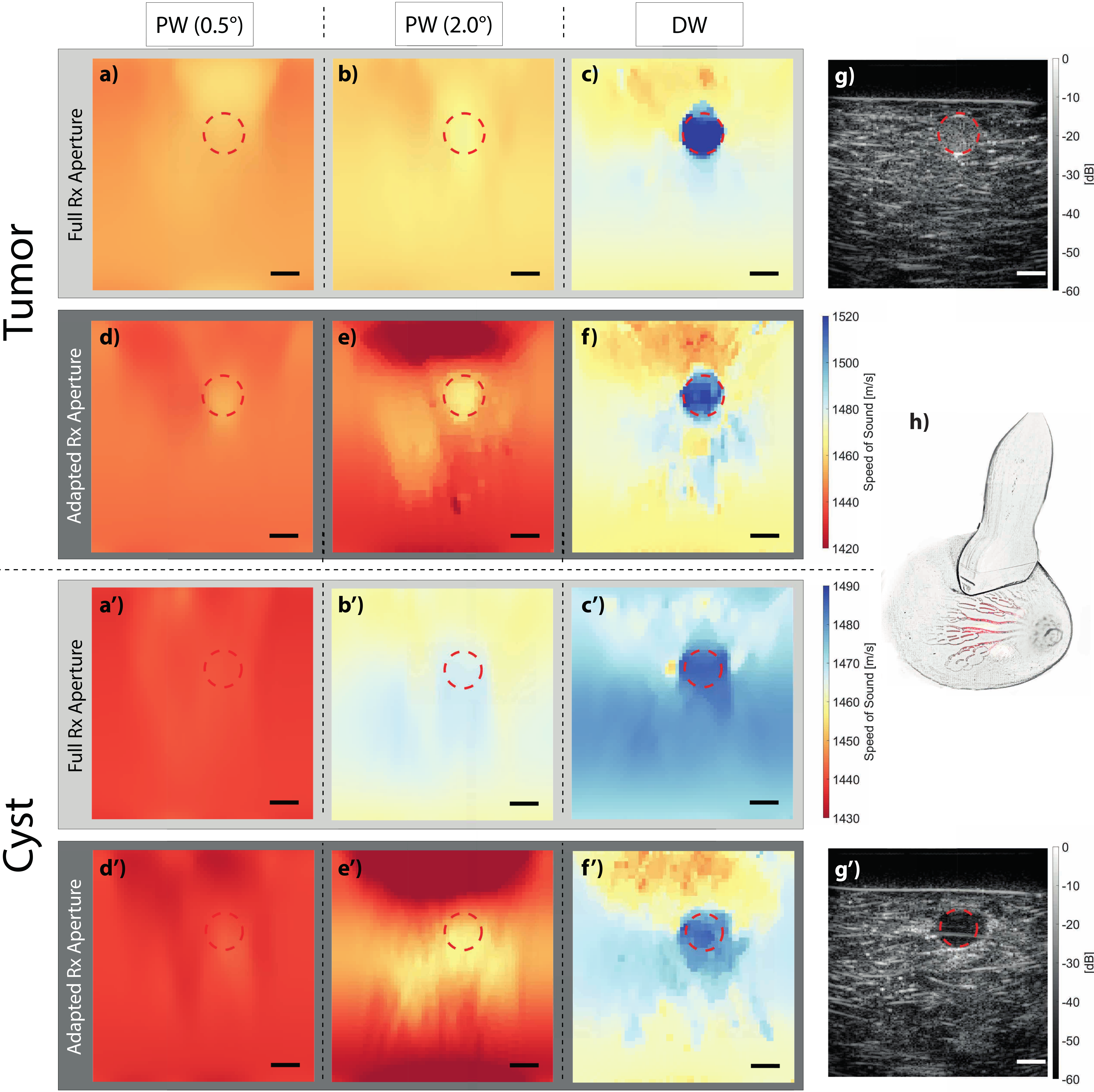}
\captionof{figure}{Reconstructions of two different cross-sections of the CIRS breast phantom.
In the upper part (a-g), a solid mass representing a malignant tumor is imaged, and in the lower part (a'-g'), a cystic structure with high echogenecity but low SoS contrast.
SoS reconstructions are shown for PW and DW with different Rx aperture approaches in (a-f,a'-f').
(g,g') show the B-Mode images with the inclusions marked, which are also overlaid in the SoS reconstructions.
(h) shows a sketch of the imaging setup with the linear array probe on the breast. 
Note that the colorbars are different for the tumor and the cyst cases. 
Scalebars represent a length of $5~\mathrm{mm}$.
}
\label{fig:phant_recons}
\end{figure}

CIRS breast phantom has stiff inclusions representing malignant solid masses with higher speed-of-sound (cf.\,Fig.~\ref{fig:phant_recons}g), and hypoechoic inclusions representing cysts (cf.\,Fig.~\ref{fig:phant_recons}g'), which have smaller SoS contrast with its surrounding. 
We reconstructed SoS maps using the optimal settings found in the previous section (cf. table~\ref{tab:FinalResults}), since we modeled this probe and acquisition scheme in our simulations. 
SoS reconstructions using the different methods are shown in Fig.~\ref{fig:phant_recons}a-f and a'-f'. 
The DW approach is seen to substantially improve the detection of both the solid mass and the cystic inclusion, whereas with PW neither the inclusion shows contrast nor the background SoS appears consistent. 
For the DW cases, the background SoS values have a higher variation, which is mainly due to the lower regularization weight ($\lambda_\mathrm{DW}$=5 vs.\ $\lambda_\mathrm{PW}$=9), indicating the inverse problem being better posed with DW.

\section{Conclusion}

We have presented herein the use of diverging waves (DW) in pulse-echo SoS image reconstruction, studying it comparatively to existing plane waves (PW) approaches.
Analyzing the wavefront aberrations with PW and DW insonifications, DW is seen to cause less aberration effects that would lead to inaccuracies in displacement estimation and thus potentially in the subsequent SoS reconstructions. 
We studied this with simulations of a set of 28 numerical phantoms and found that the quantitative accuracy ($\mathrm{RMSE}$) of SoS reconstructions is over 22\% improved by using DW instead of PW.
Even more pronounced are the improvements in inclusion contrasts, where $\mathrm{CNR}$ led to an improvement of over 55\% with DW. 
These results also translated to an actual ultrasound acquisition of a breast phantom, where the improvements with DW are qualitatively presented.
 
Diverging waves in this work are generated without loss of generality using a single element transmission yielding circular wavefronts.
Nevertheless, the formulations would easily extend to multiple-element transmit using virtual source approach and to non-circular wavefronts (i.e., by adjusting delays in beamforming and $L$ matrix paths).

With our findings SoS imaging based on conventional ultrasound systems can be substantially improved, paving the way for translating SoS imaging into the clinic. 
SoS may be a viable tool for differential diagnosis, potentially superior compared to elastogrpahy~\cite{glozman_method_2010}. 
In complex setups using ring transducers and dedicated systems~\cite{pratt_sound-speed_2007,li_breast_2017}, SoS in combination with attenuation imaging has already proven its great potential. 
For conventional US systems, it was recently also demonstrated for the first time that the attenuation distribution can be reconstructed using a passive reflector~\cite{Rau_attenuation_19}.

Quantitative SoS imaging is not only valuable as a diagnostic tool, but can also help improve other ultrasound modalities by correcting aberrations, as shown in~\cite{rau_ultrasound_2019} for beamforming of B-mode images.  
A practical limitation of our proposed SoS imaging method is that an algebraic reconstruction is utilized, which is relatively time consuming.
A variational network solution similar to~\cite{vishnevskiy_deep_2019,vishnevskiy2018im} with inference times on the order of milliseconds could help to overcome this limitation towards real-time SoS imaging.

\vspace{3ex}{\bf Funding} was provided by the Swiss National Science Foundation and Innosuisse.

\newpage

\bibliographystyle{IEEEtran}
\bibliography{refs}

\end{document}